\documentclass[12pt,a4paper]{article}
\usepackage{jheppub}

\usepackage{epsfig,multirow} 
\usepackage{amscd}
\usepackage[matrix,arrow,curve]{xy}
\usepackage{verbatim}
\usepackage{latexsym}
\usepackage{amsfonts,amsthm}

\newcommand{\Tr}{{\rm Tr}}
\newcommand{\Db}{\Bar{D}}
\renewcommand{\a}{\alpha}
\renewcommand{\b}{\beta}
\renewcommand{\l}{\lambda}
\newcommand{\intsup}{\int\!\! d^3xd^4\theta ~}  
\newcommand{\intch}{\int\!\! d^3xd^2\theta ~}

\subheader{UCSD-PTH-11-10}

\title{\begin{center} 
F-maximization along the RG flows: \\
a proposal
\end{center}}
\author[a]{Antonio Amariti,}
\author[b]{Massimo Siani}

\affiliation[a]{Department of Physics, University of California, \\
San Diego La Jolla, CA 92093-0354, USA}
\affiliation[b]{Instituut voor Theoretische Fysica, Katholieke Universiteit Leuven,\\
Celestijnenlaan 200D, B-3001 Leuven, Belgium}

\emailAdd{amariti@physics.ucsd.edu}
\emailAdd{massimo.siani@fys.kuleuven.be}

\abstract{
We propose an extension of the $F$-maximization principle to take into account the effects of non-superconformality.
Guided by a four-dimensional analog, we formulate a modification of the free energy via the Lagrange multiplier technique.
We conjecture that the Lagrange multiplier plays the same role as the coupling constant, at least at weak coupling.
We check our proposal in many examples with unitary, symplectic and orthogonal gauge groups.
}

\keywords{}

\begin{document}
\maketitle

\section{Introduction}
Rigid supersymmetry on curved backgrounds
has been recently investigated because it provides exact, i.e. all-loop and nonperturbative, results about flat space
superconformal quantum field theories. Several properties of even strongly coupled four- and three-dimensional physics were discovered with this
approach. In the case of three-dimensional field theories, the ${\rm AdS}_4 {\rm /CFT}_3$ duality provided an important motivation.
Indeed it was shown that the supergravity side of the correspondence predicts that at large $N$ the strongly coupled field theory
entropy scales as $N^{3/2}$. This result cannot be understood by using ordinary field theoretical techniques. Nevertheless,
it was shown by \cite{Drukker:2010nc} that by localizing the ABJM model on a  three sphere
the gravitational large $N$ result could be obtained even in field theory.
Then in \cite{Herzog:2010hf} this result was extended to theories with 
lower ($\mathcal{N}=3$) degree of supersymmetry. 
These computations were based on the partition function defined in \cite{Kapustin:2009kz}
with the help of the localization techniques  of  \cite{Pestun:2007rz}.

Three-dimensional $\mathcal{N}=2$ theories remained an open problem.
Indeed in this case the dimensions of the matter supermultiplets are not protected. This is due to the fact that the
$R$ symmetry is abelian and it can mix with all the other $U(1)$ global symmetries. Because the exact value
of the $R$ charges was not known, a computation along the lines of
\cite{Kapustin:2009kz} could not been performed.

The same problem of 
finding the exact $R$ charges was solved in four dimensions by maximizing the conformal anomaly coefficient
$a$ \cite{Intriligator:2003jj}. The latter appears in the trace of the stress-energy momentum tensor multiplying the Euler density.
Because in three dimensions (and more generally in odd dimensions) there are no anomalies, no analogous procedure can be
applied to three-dimensional field theories.
A former solution valid in odd and even dimensions consists in minimizing the coefficient $\tau$
in the R current two point correlation function, but this result must be quantum corrected and it
does not provide an exact answer \cite{Barnes:2005bm}.

Recently an useful alternative technique was worked out in \cite{Jafferis:2010un}.
It consists of extracting the exact superconformal $R$ charge of a three dimensional theory
by the extremization of the absolute value of the partition function
localized on $S^3$. This result followed from  a \emph{mysterious} 
holomorphy based on the supersymmetry algebra. More recently \cite{Festuccia:2011ws}
showed that this holomorphy is not mysterious but it is related to the
property of the background fields used to study the theory 
on a curved space.

This technique allows to extend the ${\cal N}=6$ \cite{Drukker:2010nc} and ${\cal N}=3$ \cite{Herzog:2010hf} 
entropy scaling behavior to many superconformal strongly coupled ${\cal N}=2$
 models, showing that it agrees with the $N^{3/2}$ dual supergravity prediction \cite{Jafferis:2011zi}.
Many different  computations  
based on \cite{Jafferis:2010un} where then performed in three dimensional 
CS matter theories.
First at large $N$ and finite CS level $k$ has been investigated in \cite{Martelli:2011qj,Cheon:2011vi,Jafferis:2011zi}.
The opposite regime was studied in \cite{Amariti:2011hw,Amariti:2011da}
by using the perturbative techniques  explained in \cite{Marino:2002fk,Aganagic:2002wv}.
Other powerful techniques in this regime 
were discussed in \cite{Minwalla:2011ma} for both large $k$ and $N$.
 At finite $N$ and $k$ some analytical computation can be implemented along the line of \cite{Dolan:2008qi,Spiridonov:2009za}
 by using the correspondence between the four dimensional superconformal index and the three 
 dimensional partition function, observed in \cite{Dolan:2011rp,Gadde:2011ia,Imamura:2011su}, while 
 in other cases numerical computation are necessary \cite{Jafferis:2011ns,Niarchos:2011sn,Minwalla:2011ma}.

All the examples discussed in \cite{Jafferis:2011zi} share the following interesting property.
When two different fixed points of the same theory are connected by an RG flow, the IR free energy $F_{IR}$ is less than
the UV one $F_{UV}$. Moreover, the free energy itself is maximized (as opposed to extremized) by the exact $R$-symmetry.
This naturally leads to conjecture that this function respects a sort of c-theorem, or equivalently, it is a good candidate for counting the
number of massless degrees of freedom, which monotonically decrease along an RG flow.

The existence of a c-theorem in QFT was first proved in  \cite{Zamolodchikov:1986gt} for the central charge
of two dimensional field theories.
In four dimensions this role is conjectured  to be played by  the central charge $a$, associated with the Euler density which
appears in the 
trace anomaly of the stress-energy tensor \cite{Cardy:1988cwa}. No counterexample to this $a$-theorem is known.
 One of the main laboratories to test and prove 
 Cardy's conjecture has been supersymmetry \cite{Anselmi:1997am,Kutasov:2003ux,Barnes:2004jj}.
Indeed it was observed that a slight generalization of the $a$-maximization procedure leads to a generalized definition of the
central charge itself, such that it can be defined even away from the fixed points \cite{Kutasov:2003ux}.
This extension relies on the addition to the conformal anomaly of a term proportional to the beta function.
This term vanishes at the fixed points, but it is multiplied by a Lagrange multiplier which, upon maximization, enforces
 the constraints from the anomaly cancellation. The conjecture made in \cite{Kutasov:2003ux} is that  
the modified anomaly coefficient $a$, when maximized with respect to the $R$ charges, provides the running $R$-symmetry
as a function of the Lagrange multiplier, and that the latter is related to the running coupling constant.
At the perturbative level this conjecture has been tested up to three
loops, where the anomalous dimensions are scheme dependent, and the opportune scheme 
relating the Lagrange multipliers and the coupling constants was discussed. Moreover
the validity of the $a$-theorem was checked in many examples, even at strong coupling \cite{Barnes:2004jj}.

As discussed above, in three dimensions  \cite{Jafferis:2011zi}  
conjectured  that there is a $c$-theorem for the free energy.
Basically the free energy takes into
account the reduction of the degrees of freedom between the extremal points of the flow,
as $c$ in two dimensions and $a$ in four.
This conjecture was then verified in some examples at weak 
 \cite{Amariti:2011da} and strong \cite{Jafferis:2011zi,Gulotta:2011si} coupling.
Moreover in \cite{Niarchos:2008jb} the study of the role of accidental symmetries in the
$F$-theorem  was started.

This $F$-theorem has not yet been formulated in stronger versions, i.e. there is no claim that the free
energy monotonically decreases along an RG flow, or that the RG flow is a gradient flow. Indeed there are not techniques to study the behavior
of the free energy away from the  fixed points.

Here, motivated by this problem, we wonder if the analogy, at the fixed point,
between $a$ in four dimensions and $F$ in three dimensions, can be extended to the whole RG flow.
We associate a Lagrange multiplier $\lambda$ to every coupling constant. At the fixed point, every $\l$ enforces the constraint
coming from the respective beta function. We check that the multipliers are related in a universal way to the running coupling constants along the RG flows in many weakly coupled examples. Thus, our modified prescription for the free energy provides an
interpolating function between the UV and IR exact results $F_{UV}$ and $F_{IR}$.

The rest of the paper is organized as follows. In section \ref{secmult} we review the conjecture about the four dimensional central charge away from the fixed points and propose the analogy
for three dimensional theories. In section \ref{sec1} we compute the anomalous dimensions by using perturbation theory
for $SU(N)_k$ gauge theories. In section \ref{sec2} we check the agreement between the perturbative computation and our proposal for the modified free energy. 
Then in section \ref{sec3} we extend the procedure to orthogonal and symplectic gauge groups, checking again the validity of our proposal. 
Finally we discuss some possible extension and the connection with the $F$-theorem.
\\
\\
{\bf Note Added}: After the appearance of our paper,  \cite{Komargodski:2011vj} appeared, where some progress on the strong version of the $a$-theorem
have been claimed. It would be interesting to match these result to the four dimensional conjecture on the Lagrange multiplier
along the RG flow, and to extend the formalism of \cite{Komargodski:2011vj}
 to three dimensions, and compare with our result. We thank the referee for suggesting 
us this connection.

\section{A proposal for the free energy  away from the  fixed point} \label{secmult}

Consider a supersymmetric UV free quantum field theory with an IR fixed point. If the theory has a $R$-symmetry, at the superconformal fixed points the $R$-current belongs
 to the same supermultiplet of the stress-energy tensor. As the theory flows according to the renormalization group equations, the $R$-symmetry can mix with all the other
 non-$R$ global abelian symmetries. Every linear combination of the UV $R$-current with the flavor abelian currents is a $R$-current. Thus, it is no longer clear how to
 identify the exact $R$-symmetry at the IR superconformal fixed point.

In the case of four-dimensional supersymmetric field theories the maximization of the coefficient of the Euler density provides the right answer. This procedure, called
 a-maximization, is only defined at the fixed point, thus it does not give any information about the $R$-symmetry away from the fixed point itself. Nevertheless, it was
 conjectured that a slight generalization  can be used to find the running $R$-current which interpolates between the exact UV and IR ones \cite{Kutasov:2003ux}. 
This idea has been used to study a stronger version of the a-theorem, along the RG flow \cite{Barnes:2004jj}.

Usually both gauge interactions and superpotential couplings change during the flow. In three-dimensional Chern-Simons-matter theories
there is no continuous gauge coupling and we discuss only the RG running of the superpotential couplings, bearing in mind that an analogous 
discussion  for the gauge couplings holds in four dimensions.

Let us first review the four-dimensional construction. 
The aim is defining  a running central charge $a$ which interpolates between the UV and the IR ones. At the UV fixed point, 
the coupling constants in the superpotential
\begin{equation} \label{SPOTUV}
W = h_{ijk} Q_i Q_j Q_k
\end{equation}
vanishes. At the IR fixed point, the $R$-charges of the fields are found by maximizing the central charge after  the marginality of the superpotential
(\ref{SPOTUV}) is imposed.
 The latter constraints can be imposed by adding a Lagrange multiplier to the central charge, that is by defining
\begin{equation}
a(\lambda,R) = a(R) + \lambda_h ( R_i+R_j+R_k-2) 
\end{equation}
where  $\beta_{ijk} \simeq  \tilde \beta_{-ijk} \equiv  R_i+R_j+R_k-2$.
The Lagrange multiplier $\l_h$ and the coupling constant $h$ are related by \cite{Kutasov:2003ux,Barnes:2004jj}
\begin{equation} \label{rellambda}
\lambda_h = \frac{h^*_{ijk} h^{ijk}}{24 \pi^2} +\dots = \frac{
|h|^2 T^*_{ijk} T^{ijk}}{24 \pi^2}+\dots= \frac{|h|^2 |T|^2}{24 \pi^2} +\dots
\end{equation}
where $\dots$ refers to higher loop scheme dependent corrections and $T_{ijk}$ is the group-theoretical structure of the interaction.
This conjecture was supported by a plenty of examples, in which it was shown that at weak coupling the R charge $R(\lambda_h)$ matches with the perturbative $R(h_{ijk})$
 even away from the fixed point.

Here we propose that a similar relation holds in three-dimensional SCFTs. In this case the number of massless degrees of freedom is believed to be encoded in the free
energy $F$, which can be computed from the partition function $|{\cal Z}|=e^{-F} $ localized on $S^3$. We will give evidences that the following modification
of the partition function
\begin{equation}
|\mathcal{Z}(R,\lambda)| = e^{-F(R) + \lambda_i \beta_i(R) \log |\mathcal{I}_k| }
\label{eq:proposal}
\end{equation}
takes into account the effects of the non-conformality. In the formula above,
\begin{equation}
\mathcal{I}_k = \int_{G} \prod_{i=1}^{\text{rank}\, G} \text{d} u_i \prod_{i<j} 2 \sinh^2(\pi(u_i-u_j)) e^{i k \pi \text{Tr}_{F} u^2}
\end{equation}
In the perturbative regime  the superpotential flow is usually generated by
the terms 
\begin{equation}
W = h_{ijkl} \phi_i \phi_j \phi_k \phi_l
\end{equation}
and  between $h_i$ and $\lambda_i$  we propose the relation
\begin{equation}
\lambda_h = \frac{|h|^2 |T|^2}{32} +\dots
\label{eq:relation}
\end{equation}
The equations (\ref{eq:proposal}) and (\ref{eq:relation}) constitute our main results.
In the rest of the paper we verify this statement at the leading order in various weakly coupled examples.

We will not address two problems that we leave for future investigations.
The first one is the scheme dependence of the multiplier. Indeed the leading order two loops computation is scheme independent.
It is important to understand whether the scheme independent part of higher loop computations and that of higher order expansions of the partition function match.
This would also fix the scheme dependent coefficients.
The second important question deals with the strongest version of the $c$-theorem, i.e. the RG flow is a gradient flow of the $c$-function.
Indeed in four dimensions it was observed that  the relation
\begin{equation}
\partial_{\lambda_I} a(\lambda) =\tilde \beta^I(R(\lambda))  
\end{equation}
suggests that the RG flow is a gradient flow.  
We can parametrize the $\beta$ functions of the theory as
\begin{equation}
\tilde \beta^I(R(\lambda)) = f_J^I (g) \beta^{J}(g)
\end{equation}
thus
\begin{equation}
\partial_{h_I} a = \partial_{\l_K} a \,\, \partial_{h_I} \l_K = \partial_{h_I} \l_K \,\, f_J^K \beta^{J}
\end{equation}
The matrix $G_{IJ}\equiv \partial_{h_I} \l_K \,\, f_J^K$ provides the metric of the coupling space, and the 
$F$-theorem requires its positivity.

\section{Field theory results} \label{sec1}

In this section we present the field theory results that we will use in the following sections. We will give the explicit relation between the anomalous
 dimensions and the coupling constants, and their values at the fixed points. This will allow us to compare the field theory computations with the ${\cal Z}$
 extremization procedure to give evidence for our proposal about the form for the partition function outside the fixed point.

We always consider a $\mathcal{N}=2$ gauge theory with a Chern-Simons term.
The vector multiplet $V$ is in the adjoint representation of the gauge group, and it is eventually coupled to some matter multiplet in some representation
 of the local and global symmetry. We will be mainly concentrating on the $SU(N)$ gauge group, but we will also comment on the orthogonal and symplectic
 cases. Thus, in the superspace language the gauge sector action reads
\begin{equation}
{\cal S}_{\mathrm{CS}}
  = \frac{k}{4\pi} \int d^3x\,d^4\theta \int_0^1 dt\: \Tr \Big[
  V \Db^\a \left( e^{-t V} D_\a e^{t V} \right) \Big]
  \label{eq:actionCS} 
\end{equation}
where $k$ sets the Chern-Simons level, which is an integer number for the theory to be invariant under large gauge transformations.

It would be interesting to check our proposal to the flavor deformations of the ABJM model. 
While deformations of the bifundamental sector \cite{Akerblom:2009gx} only lead to finite theories at the fixed
 points \cite{Bianchi:2009ja,Bianchi:2009rf}, adding flavors provides a nontrivial spectrum of non finite fixed points \cite{Bianchi:2009rf,Amariti:2011hw}.

\subsection{Fundamentals and singlets}

The simplest nontrivial model that we will use to test our proposal (\ref{eq:proposal}) contains $N_f$ pairs of fields $q_a^r, \tilde q_r^a$, $a=1,\ldots,N, r=1,\ldots,N_f$
 in the (anti)fundamental representation of the gauge group $SU(N)$. Adding a superpotential deformation $W \sim (q \tilde q)^2$ would fix their $R$-charge at the fixed
 point to the free field value $R=\frac{1}{2}$. To avoid this, we couple them to $|G_f|$ gauge singlet fields $M_{rs}\equiv M^A (T_f^A)_{rs}$ in the adjoint representation
 of the flavor group $G_f$. 
Thus, we are led to consider the action
\begin{equation}
\begin{split}
{\cal S} = {\cal S}_{\rm CS} &+ \intsup \Tr \left( {\bar q}_r e^V q^r +
  \bar{\tilde{q}}^r \tilde q_r e^{-V} \right) + \intsup \bar M M \\ &+ \left( \intch W + {\rm h.c.} \right)
\end{split}
\label{eq:fundsing}
\end{equation}
with
\begin{equation}
W = h \, \Tr \tilde q M^2 q
\label{eq:Wfundsing}
\end{equation}
The trace over the flavor group in the singlets kinetic term is understood. The UV deformation (\ref{eq:Wfundsing}) is the only one which does not fix any $R$-charges
 at the fixed point to acquire the value $R=\frac{1}{2}$. At the lowest nontrivial order the fundamental fields get quantum corrections both from the coupling with the
 gauge field and from the superpotential, while the leading order singlets anomalous dimensions only depend on the latter coupling. We choose the global symmetry to be
 $SU(N_f)$ and carry out the two loop computation which gives
\begin{equation}
\begin{split}
\gamma_q &= -\frac{N^2-1}{2 k^2 N^2}(N N_f-1) + \frac{|h|^2}{32\pi^2} \frac{N_f^2-1}{N_f^2} (N_f^2-2) \\
\gamma_M &= \frac{|h|^2}{16\pi^2} \frac{N}{N_f} (N_f^2-2) \\
\end{split}
\label{eq:gammaUNf}
\end{equation}
At the IR fixed point $\l\neq 0$ we impose the anomaly cancellation constraint $\b_\l \propto \gamma_q+\gamma_M=0$ and extract the coupling constant value
\begin{equation}
|h^\ast|^2 = \frac{16 N_f^2 (N_f N-1) \left(N^2-1\right) \pi^2}{k^2 N^2 (N_f^2-2) (N_f^2+2 N N_f-1)}
\label{eq:lambdaU}
\end{equation}
We will compare the equations (\ref{eq:gammaUNf}) and (\ref{eq:lambdaU}) with the same quantities computed from the partition function to check that our proposal
 correctly reproduces the $\l$ dependence of the anomalous dimensions. 
Even if the factors in (\ref{eq:gammaUNf}) are simply related to Casimir operators,
  it is nontrivial for  the gauge and flavor group factors to satisfy  (\ref{eq:relation}).

\subsection{Fundamentals and adjoints}
In our next example all the fields will be charged under the gauge group.
In addition to the $N_f$ pairs of quarks $q_a^r$, $\tilde q_r^a$ described above we introduce in the theory $g$ ${\cal N}=2$
chiral fields $\phi_i$, $i=1,\ldots,g$ in the adjoint representation of the gauge group $G$, namely $\phi_i \equiv \phi_i^A T^A$
 where $A=1,\ldots,|G|$ and $T^A$ are the generators of $G$. Adding
\begin{equation}
\intsup \Tr \left( \bar \phi^i e^V \phi_i e^{-V} \right)
\end{equation}
and the superpotential couplings
\begin{equation}
 W = \sum_{i=1}^{g} \alpha_1 \phi_i^4 + \alpha_2 q \phi_i^2 \tilde q + \alpha_3 (q \tilde q)^2
 \label{eq:W1group}
\end{equation}
to (\ref{eq:actionCS}) and the fundamental sector we obtain \cite{Amariti:2011da}
\begin{equation}
\begin{split}
\gamma_\phi &= -\frac{N(N+N_f+g N)}{k^2} + \frac{1}{32\pi^2} \left[ 2 \left|\a_2\right|^2 N_f \frac{N^2-2}{N} + 16 \left|\a_1\right|^2 {\cal J} \right] \\
\gamma_q &= -\frac{\left(N^2-1\right) (N(N_f+g N)-1)}{2 k^2 N^2} \\
            & \:\:\: + \frac{1}{32\pi^2} \left[ \left|\a_2\right|^2  g \frac{N^2-1}{N^2} \left( N^2-2 \right) + 4 \left|\a_3\right|^2 \left(N N_f+1 \right) \right]
\end{split}
\end{equation}
where $\cal J$ is an irrelevant group-theoretical factor for our purposes.
Considering a non-vanishing $\a_1$ or $\a_3$ leads to $R_\phi=\frac{1}{2}$ or $R_q=\frac{1}{2}$, respectively. Thus, we set
 $\a_1=\a_3=0$ and consider the fixed point value of the coupling $\a_2$
\begin{equation}
\left| \a_2 \right|^2 = \frac{16 \left(1-(1+g) N^2+(2+3 g) N^4+ N_f N \left(3 N^2-1\right)\right) \pi^2}{k^2 \left(N^2-2\right) \left(2 N_f N+g \left(N^2-1\right)\right)}
\end{equation}

\subsection{General ${\cal N}=2$ adjoint models}
We now consider the Chern-Simons-matter theories with only adjoint fields as in 
\cite{Gaiotto:2007qi}.

We consider a relevant superpotential deformation  
\begin{equation}  \label{spotculo}
W = h_{ijkl} \Tr \, \phi_i \phi_j \phi_k \phi_l
\end{equation}
We discuss 
three different possibilities $h_{1122}\neq0$, $h_{1123} \neq 0$ and
$h_{1234}\neq 0$. For simplicity we study in every
case the situation in which the adjoint fields involved in the superpotential
represents all the matter fields. It follows that there are two adjoint fields in the first case, 
three in the second and four in the last.
The universal gauge contribution, which only depends on the total number $g$ of fields, is
\begin{equation} \label{anomal}
\gamma_{\phi_i} = -\frac{N^2 (g+1)}{k^2}
\end{equation}
By adding the superpotential (\ref{spotculo}) we have different RG flow.  

\subsubsection*{\underline {$h=h_{1122}$}}
This case corresponds to $W=h \, \Tr \, \phi_1^2 \phi_2^2$.  Here $g=2$ and the anomalous dimensions for the adjoint fields
are shifted from (\ref{anomal}) by
\begin{equation}
\delta \gamma_{\phi_1}=\delta \gamma_{\phi_2}=  \frac{|h|^2}{16 \pi^2 N^2} (N^4-4N^2+12) 
\end{equation}
At the fixed point the coupling constant is
\begin{equation}
|h^\ast|^2 = \frac{48 \pi^2 N^4 }{k^2  (N^4-4N^2+12)}
\end{equation}
\subsubsection*{\underline {$h=h_{1123}$}}
This case corresponds to $W=h \, \Tr \, \phi_1^2 \phi_2 \phi_3$. Here $g=3$ and the anomalous dimensions for the adjoint fields
are shifted from (\ref{anomal}) by
\begin{eqnarray}
\delta \gamma_{\phi_1}&=&      \frac{|h|^2}{16 \pi^2} \frac{N^4-4N^2+6}{N^2}      \\
\delta \gamma_{\phi_2}&=&\delta \gamma_{\phi_3}=  \frac{|h|^2}{32\pi^2} \frac{N^4-4N^2+6}{N^2}  \nonumber 
\end{eqnarray}
At the fixed point the coupling constant is
\begin{equation}
|h^\ast|^2 =\frac{256 \pi^2 N^4}{3 k^2 (N^4-4N^2+6)}
\end{equation}
\subsubsection*{\underline {$h=h_{1234}$}}
This case corresponds to $W=h \, \Tr \, \phi_1 \phi_2 \phi_3 \phi_4$. Here $g=4$ and the anomalous dimensions for the adjoint fields
are shifted from (\ref{anomal}) by
\begin{equation}
\delta \gamma_{\phi_i}= \frac{|h|^2}{32 \pi^2} \frac{N^4-3N^2+3}{N^2} 
\end{equation}
At the fixed point the coupling constant is
\begin{equation}
|h^\ast|^2=\frac{160 \pi^2 N^4}{k^2 (N^4-3N^2+3)}
\end{equation}

\section{Extremization with Lagrange multipliers} \label{sec2}

In this section we reproduce the perturbative field theory results obtained
above by adding a Lagrange multiplier to the free energy as described in section \ref{secmult}.
The modified free energy $F(\Delta,\lambda)$, is a function of both  $\Delta$, the conformal
dimensions of the fields,  that in three dimensions corresponds  to the $R$ symmetry and
$\lambda$, the Lagrange multiplier that we have conjectured to be related with the coupling constant.
The extremization of $F$ with respect to $R$
\begin{equation}
\partial_{\Delta} F(\Delta, \lambda)=0
\label{eq:extremization}
\end{equation}
gives the behavior of the $R$ charge as a function of the Lagrange multiplier $\lambda$. The anomalous dimensions can be extracted from
\begin{equation}
\Delta(\lambda)= \frac{1}{2} + \gamma(\lambda)
\end{equation}
We conjecture that the Lagrange multiplier is related to the coupling constants, so that (\ref{eq:extremization}) and (\ref{eq:relation}) give the correct anomalous dimensions as
 a function of the couplings.

When we insert the $R$ charges back in the free energy and extremize with respect to $\lambda$ 
\begin{equation}
\frac{\text{d} F(\lambda,\Delta(\lambda))}{d \lambda} = 0 
\end{equation}
we find the value $\lambda^*$ of the multiplier at the fixed point.
With this strategy we obtain both the $R$ charges (or the anomalous dimensions) and the Lagrange multiplier at the fixed point.
 Then we verify  that the relation (\ref{eq:relation})
is valid, relating the Lagrange multiplier and the coupling constant.

\subsection{$SU(N)_k$ theories with $N_f$ fundamentals and  singlets}

The first model that we discuss is the $SU(N)_k$ YM-CS
gauge theory, at large $k$ and finite $N$ with $N_f$ fundamentals $q$ and $\tilde q$ and 
a singlet $M$. 
The UV theory is deformed by the relevant superpotential (\ref{eq:Wfundsing}), and the RG flow ends 
in a weakly coupled IR fixed point. The anomalous dimensions of $q$, $\tilde q$ and $M$ 
can be computed from the extremization of the absolute value of the partition function
\begin{equation}
\mathcal{Z} = \int d[u] \Delta\left(2\sinh^2 (\pi u) \right) e^{i k \pi \text{Tr}_F u^2 }
e^{(N_f^2-1) l(1-\Delta_M) } \prod_{i=1}^{N_c} e^{N_f l(1-\Delta_{q} -i u_i) + N_f l(1-\Delta_q +i u_i) }
\end{equation}
with the constraint $\Delta_M+\Delta_q=1$ imposed by the superpotential.
As explained above the constraint can be imposed at a dynamical level on the free energy,
$F=-\log|\mathcal{Z}|$.
From (\ref{eq:proposal}) we have 
\begin{equation} \label{freeatall} 
F (\Delta_M,\Delta_q,\lambda_h) = F(\Delta_M,\Delta_q)+\lambda_h (\Delta_M+\Delta_q-1)|\mathcal{I}_k|
\end{equation}

We computed the partition function as explained in \cite{Amariti:2011hw,Amariti:2011da} at the second order in the 't Hooft coupling
and extremized the free energy (\ref{freeatall})  with respect of the $R$ charges obtaining
\begin{eqnarray} \label{scemo2}
\Delta_q &=& \frac{1}{2}+\gamma_q(\lambda_h)=\frac{1}{2}- \frac{(N^2-1)(N_f N-1)}{2 k^2}+\frac{2 \lambda_h}{N N_f \pi^2} \nonumber \\
\Delta_M &=& \frac{1}{2}+\gamma_M(\lambda_h)= \frac{1}{2}+ \frac{2 \l_h}{(N_f^2-1) \pi^2} 
\end{eqnarray}
By substituting (\ref{scemo2}) in (\ref{freeatall}) we have $F=F(\lambda)$. The maximization of this quantity 
gives the value of the Lagrange multiplier at the fixed point
\begin{equation}
\l_h^* = \frac{ N_f (N_f^2-1) (N N_f-1) (N^2-1) \pi^2}{2 N k^2 (N_f^2+2 N_f N-1)}
\end{equation}
which as expected is related to the coupling constant by (\ref{eq:relation}). Indeed in this case we expect
\begin{equation}\label{fixed}
\l_h^*= \frac{|h^*|^2}{32} |G_f| |r_G| \left( 2 C_2(N_f) - \frac{1}{2} C_2(G_f) \right)
\end{equation}
where $|r^q_G|=N$ for $G=SU(N)$ and $q$ in the fundamental representation.
Explicitly we have
\begin{equation}  \label{culo}
\l_h^* = \frac{|h^\ast|^2}{32}\frac{ N (N_f^2-1) (N_f^2-2)}{N_f}
\end{equation}
which  is satisfied by (\ref{eq:lambdaU}) and (\ref{fixed}). Moreover (\ref{culo}) is not only respected at the fixed point but even during the flow,
as can be observed by comparing (\ref{eq:gammaUNf}) and  (\ref{scemo2}).

\subsection{$SU(N)_k$ theories with $N_f$ fundamentals and  $g$ adjoints}
The second model  is a $SU(N)_k$ YM-CS
gauge theory, at large $k$ and finite $N$ with $N_f$ fundamentals $q$ and $\tilde q$ and 
$g$ adjoints $\phi$. 
We study the flow generated by the superpotential (\ref{eq:W1group}) with $\alpha_1=\alpha_3=0$. 
The partition function for this model has been studied in \cite{Amariti:2011da}. By extremizing the free energy $F(\Delta_\phi,\Delta_q,\lambda_{\alpha_2})$ to respect of
$\Delta$  we obtain the $R$ charges during the flow
\begin{eqnarray}\label{cicciopidocchio}
R_{q}&=& \frac{1}{2}+\gamma_q(\lambda_{\alpha_2}) = \frac{1}{2} -\frac{(N^2-1)(N(N_f+ g N)-1)}{2 k^2 N^2}+\frac{\lambda_{\alpha_2}}{N_f N \pi^2}
\nonumber \\
R_{\phi} &=& \frac{1}{2}+\gamma_\phi(\lambda_{\alpha_2}) = \frac{1}{2} -\frac{N(N_f+N(g+1))}{k^2}+\frac{2 \lambda_{\alpha_2}}{g(N^2-1)\pi^2}
\end{eqnarray}
Inserting (\ref{cicciopidocchio}) in $F(\Delta_\phi,\Delta_q,\lambda_{\alpha_2})$
the Lagrange multiplier at the fixed point is
\begin{equation}
\lambda_{\alpha_2}^* = \frac{g N_f(N^2-1)(1-(g+1)N^2+(2+3 g)N^4+N_f N(3N^2-1) \pi^2}{2 N(2 N_f N+g(N^2-1)) k^2}
\end{equation}
In this case the expected relation between $\lambda_{\alpha_2^*}$ and 
$|\alpha_2^*|$ is 
\begin{equation}
\l_h^*= \frac{|h^*|^2}{32} |G| N_f g \left( 2 C_2(N) - \frac{1}{2} C_2(G) \right)
\end{equation}
and we observe that it is 
valid all along the RG flow. Moreover we explicitly checked the agreement of the proposal for the other
more simple cases $\alpha_1 \neq 0$ and $\alpha_3\neq 0$.

\subsection{$SU(N)_k$ with multiple adjoint matter fields}

The last model is a $SU(N)_k$ YM-CS
gauge theory, at large $k$ and finite $N$ with multiple 
adjoint fields $\phi$. 

We deform this model with a superpotential (\ref{spotculo}) and discuss
three different possibilities $h_{1122}\neq0$, $h_{1123} \neq 0$ and
$h_{1234}\neq 0$. For simplicity we study in every
case the situation in which the adjoint fields involved in the superpotential
represent all the matter fields. It follows that there are two adjoints in the first case, 
three in the second and four in the last.
The partition function for the case of multiple adjoint fields have been studied in 
\cite{Amariti:2011hw} and here the only difference is the presence of the Lagrange multiplier in the 
free energy.
The anomalous dimensions $\gamma_\lambda$ and $\lambda_h^*$ are summarized in the following table.
\begin{center}
\begin{tabular}{c||c|c} 
Coupling        &     $ \gamma(\lambda_h) $  & $\lambda^{*}$ \\
\hline
$h_{1122}$    &$\gamma_1=\gamma_2=-\frac{3 N^2}{k^2}+ \frac{2 \lambda}{(N^2-1) \pi^2}$& $ \frac{3}{2}N^2(N^2-1)\pi^2 $\\
\hline
$h_{1123}$     &
$
\begin{array}{c}
\gamma_1= -\frac{4 N^2}{k^2}+\frac{4 \lambda }{(N^2-1) \pi^2}   \\
 \hline
\gamma_2=\gamma_3= -\frac{4 N^2}{k^2}+\frac{2 \lambda }{(N^2-1) \pi^2} \nonumber
\end{array}
$
&
$\frac{4}{3} N^2(N^2-1) \pi^2$
 \\
\hline
$h_{1234}$    &$\gamma_i=-\frac{5 N^2}{k^2}+\frac{2 \lambda}{(N^2-1) \pi^2} $& 
$\frac{5}{2} N^2(N^2-1) \pi^2$
\\
\end{tabular}
\end{center}

Even in that case  we observe that the expected
relation between the coupling constant and the multiplier holds during the RG flow.

\section{Orthogonal and symplectic gauge groups} \label{sec3}

In this section we discuss the relation between RG flow
of the UV relevant superpotential couplings and the 
Lagrange multiplier with orthogonal and symplectic gauge groups.
 In all the cases the flow is generated by an $SU(N_f)$ invariant superpotential
coupling the fundamental with a singlet $M$
\begin{equation} \label{SOSPspot}
W = M^2 q \cdot \tilde q
\end{equation}
where the product $q \cdot \tilde q$ is properly chosen for the  $SP(2 N)_{2k}$ and $SO(N)_{2k}$ cases
and $q=(N,\overline N_f)$, $\tilde q=(N,  N_f)$.
The Casimir for the fundamental and the adjoint representation of these groups are
\begin{center}
\begin{tabular}{c||c|c|c|c}
Group & Rep.& $|r|$ &      T(r)&$C_2(r)$\\
\hline
\hline
SU(N) &Fund.& $N$           &  $1$   &$ \frac{(N^2-1)}{ N }$\\
SU(N) & Adj.&$N^2-1$     &  $2N$&  $2N$         \\
\hline
SO(N) &Fund.& $N$              &  $2$   &$ N-1$\\
SO(N) & Adj.&$\frac{N(N-1)}{2}$     &  $2(N-2)$&  $2(N-2)$         \\
\hline
SP(2N) &Fund.& $2N$&$1$&   $ \frac{2N+1}{2}   $\\
SP(2N)   & Adj.&$\frac{2N(2N+1)}{2}$&$2(N+1)$& $2(N+1)$ \\
\end{tabular}
\end{center}
where $C_2(r) = \frac{|G|}{|r|} T(r)$. 
Recall the general formula
\begin{equation}
\gamma_q = -\frac{1}{2 k^2 T(N)^2} C_2(N) \left[ C_2(N) - \frac{C_2(G)}{2} + \frac{T(N)}{2} N_f^\prime \right]
\end{equation}
for the gauge contributions. $N_f^\prime$ is the total number of fundamental fields $N_f^\prime=2N_f$.
In the case of $SO(N_c)$ gauge symmetry with a $SU(N_f)$ global symmetry,
the two loops anomalous dimensions are
\begin{eqnarray}
\gamma_q &=& -\frac{1}{8 k^2} (N-1) (2 N_f+1) + \frac{|h|^2}{32 \pi^2 N_f^2} (N_f^2-1)(N_f^2-2)  \quad \quad \text{and} \\
\gamma_M &=& \frac{|h|^2 N}{16  \pi^2N_f} (N_f^2-2)
\end{eqnarray}
and at the fixed point the coupling is
\begin{equation}
|h^\ast|^2 = \frac{4 \pi^2 N_f^2 (N-1) (2N_f+1) }{k^2  (N_f^2-2) ( N_f^2+2 N_f N-1)}
\end{equation}
For the symplectic case 
\begin{eqnarray}
\gamma_q &=& -\frac{2N+1}{8 k^2} (2N_f-1) + \frac{|h|^2}{32 \pi^2 N_f^2} (N_f^2-1)(N_f^2-2) \quad \quad \text{and} \\
\gamma_M &=& \frac{|h|^2 N}{8\pi^2 N_f} (N_f^2-2) 
\end{eqnarray}
and at the fixed point the coupling is
\begin{equation}
|h^\ast|^2 = \frac{4 \pi^2 N_f^2 (2N+1) (2N_f-1)  }{k^2 (N_f^2-2) (N_f^2 + 4 N N_f-1)}
\end{equation}

\subsection{Partition function for SO/SP}

We compute the partition function for the $SO(N)_{2k}$ and $SP(2N)_{2k}$ gauge groups.
Analogous  computations of $\mathcal{Z}$ for symplectic gauge groups were performed in \cite{Willett:2011gp}
while the orthogonal case was studied in \cite{Kapustin:2011gh}.
The simple roots, i.e. the weights for the adjoint representation,  are, as usual, read from the Dynkin diagram. 
The $SP(2N)$ case corresponds 
to $C_N$, while the $SO(N)$ case is split in $SO(2N+1)$ corresponding to $B_N$ 
and $SO(2N)$ corresponding to $D_N$.
We then couple the fundamental fields with a $SU(N_f)$ adjoint $M$, with the superpotential
(\ref{SOSPspot}).
The partition function 
is
\begin{equation}
\mathcal{Z}= e^{(N_f^2-1)l(1-\Delta_M)} \int d[u] e^{2 i \pi k \text{Tr}_F u^2} F_{v}^{(1)} F_{m}^{(1)}(\Delta_q)
\end{equation}
and the 1-loop determinants for the vector and matter multiplet become 
\begin{itemize}
\item$SP(2N)$
\begin{eqnarray}
F_{v}^{(1)} &=&
\prod_{i=1}^{N} 2 \sinh^2 (2 \pi u_i) \prod_{i<j}\left(4 \sinh(\pi(u_i-u_j) )\sinh(\pi(u_i+u_j))\right)^2 \nonumber \\
F_{m}^{(1)}(\Delta_q) &=&\prod_{i=1}^{N}\prod_{\eta=\pm 1}  e^{2N_f l(1-\Delta_q + i \eta  u_i)}
\end{eqnarray}
\item $SO(2N)$
\begin{eqnarray}
F_{v}^{(1)} &=&\prod_{i<j}\left(4 \sinh(\pi(u_i-u_j) )\sinh(\pi(u_i+u_j))\right)^2 \nonumber\\
F_{m}^{(1)}(\Delta_q) &=& \prod_{i=1}^{N}\prod_{\eta=\pm 1} e^{2N_f l(1-\Delta_q + i \eta  u_i)}
\end{eqnarray}
\item$SO(2N+1)$
\begin{eqnarray}
F_{v}^{(1)} &=&
\prod_{i=1}^{N}  2 \sinh^2 (\pi u_i)\prod_{i<j}\left(4 \sinh(\pi(u_i-u_j) )\sinh(\pi(u_i+u_j))\right)^2 \nonumber \\
F_{m}^{(1)}(\Delta_q) &=& e^{2N_f l(1-\Delta_q) }\prod_{i=1}^{N}\prod_{\eta=\pm 1} e^{2N_f l(1-\Delta_q + i \eta  u_i)}
\end{eqnarray}
\end{itemize}
where in every case $i,j=1,\dots N$.  The factor $2 N_f$ in the one loop determinant of the matter fields 
is related to the $SU(N_f)$ flavor symmetry which distinguishes the $q$ and $\tilde q$ fields.

After adding to the free energy, computed from the partition function, the Lagrange multiplier, which imposes the constraint $\Delta_M+\Delta_q=1$,
we computed the  anomalous dimension during the flow as a function of the multiplier itself. 
We summarize the anomalous dimensions 
during the RG flow and the fixed point value of the multiplier
in the table.
\begin{center}
\large
\begin{tabular}{c||c|c}
                  & $\gamma_{q} (\lambda)$ & $\lambda^*$ \\
\hline
$SO(4) $  & $-\frac{3(2N_f+1)}{8k^2}+\frac{ \lambda}{4 N_f \pi^2 } $&
$\frac{3 N_f (2 N_f+1) (N_f^2-1) \pi^2}{2 (N_f(N_f+8)-1) k^2} $\\
\hline
$SO(5) $  & $-\frac{2N_f+1}{2 k^2}+\frac{\lambda}{5 N_f \pi^2 } $&
$\frac{5 N_f (2 N_f+1) (N_f^2-1) \pi^2}{2 (N_f (N_f+10)-1) k^2} $\\
\hline
$SO(6) $  & $-\frac{5(2N_f+1)}{8k^2}+\frac{ \lambda}{6 N_f \pi^2 } $&
$\frac{15 N_f (2 N_f+1) (N_f^2-1) \pi^2}{4 (N_f(N_f+12)-1) k^2} $\\
\hline
$SP(2) $ &  $-\frac{3(2 N_f-1)}{8 k^2}+\frac{\lambda}{2 N_f \pi^2 } $&
$\frac{3 N_f (2N_f-1) (N_f^2-1) \pi^2}{4 (N_f(N_f+4) -1)k^2} $\\
\hline
$SP(4) $ &  $-\frac{5(2 N_f-1)}{8k^2}+\frac{\lambda}{4 N_f \pi^2 } $&
$\frac{5 N_f (2N_f-1) (N_f^2-1) \pi^2}{2 (N_f (N_f+8) -1)k^2} $\\
\hline
$SP(6) $ &  $-\frac{7(2 N_f-1)}{8 k^2}+\frac{\lambda}{6 N_f \pi^2 } $&
$\frac{21 N_f (2N_f-1) (N_f^2-1) \pi^2}{4 (N_f (N_f+12)-1) k^2} $\\
\end{tabular}
\end{center}
\normalsize
We expect a general relation at arbitrary $N$ given by
\begin{equation}
\begin{split}
\gamma_q^{SO(N)} &= -\frac{(N-1) (2N_f+1)}{8k^2}+\frac{ \lambda}{N N_f \pi^2 } \\
\l^{\ast SO(N)} &= \frac{N (N-1) \, N_f (2N_f+1) (N_f^2-1) \pi^2}{8 k^2 (N_f^2+2 N N_f -1)} \\
\gamma_q^{SP(2N)} &= -\frac{(2N+1) (2N_f-1)}{8k^2}+\frac{ \lambda}{2 N N_f \pi^2 } \\
\l^{\ast SP(2N)} &= \frac{N (2N+1) \, N_f (2N_f-1) (N_f^2-1) \pi^2}{4 k^2 (N_f^2+4 N N_f -1)}
\end{split}
\end{equation}
In all the cases the $\lambda$ dependence of $\gamma_M$ is independent from the gauge group
because it is a singlet.   Explicitly we have
\begin{equation}
\gamma_{M}(\lambda) = \frac{2 \lambda}{(N_f^2-1) \pi^2}
\end{equation}
Even in this case the relation (\ref{culo}), between the coupling constant and the Lagrange multiplier,
is satisfied, not only at the fixed point, but also during the flow.

\section{Conclusions and discussion}

In this paper we conjectured that the RG flow among superconformal fixed points of 
a three dimensional field theory can be followed with the help of the extremization of the free energy localized on $S^3$. This idea is strongly connected with the analogy between
the four dimensional conformal anomaly and the three dimensional free energy \cite{Jafferis:2011zi}.
As in \cite{Kutasov:2003ux} the basic observation is that away from the  fixed point the beta function for the coupling constant does not vanish, while at the fixed point 
$\beta=0.$ This suggests that by adding a Lagrange multiplier to the free energy it becomes possible to compute the 
$R$-charges as a function of the multiplier. This procedure gives a  
relation between the coupling constant and the  Lagrange multipliers at the fixed point. The conjecture is that relation
can be extended even during the flow  away from the  fixed point. 
Usually this relation is scheme dependent, a priori it is not known which is the trajectory followed by the multiplier during the RG flow. 
Anyway in this paper we skipped over this problem because we only checked that the relation between the Lagrange
 multiplier and the coupling constant is valid at two loops, where there is no scheme dependence.
We believe that this paper is a first step towards the formulation of a stronger version of the
$F$-theorem in three dimensional field theories.

Many problems and open questions then arise.
As discussed above the first problem is the scheme dependence of the Lagrange multipliers. This requires a higher loop computation and a higher 
order expansion of the partition function.
Higher order computation are also necessary to check a stronger version of the $F$ theorem at perturbative level. Indeed here we just
observed that at two loops the Lagrange multiplier captures the quantum effects. Anyway the two loops analysis is 
not enough, as explained in \cite{Amariti:2011da}, for checking the $F$-theorem in superpotential RG flows, and a four loops computation is necessary.
Connected with this problem we observe that at two loop in perturbation theory there 
are other possible ways to add a Lagrange multiplier, to Z instead then to F
\begin{equation}
\mathcal{Z}(\Delta,\lambda) = \int_{G} \prod_{i=1}^{\text{rank}\, G} \text{d} u_i \prod_{i<j} 2 \sinh^2(\pi(u_i-u_j)) e^{i k \pi \text{Tr}_{F} u^2} \prod_{\rho}
e^{l(1-\Delta+ i \rho_I (u))}  e^{\lambda^W \beta_W}
\label{eq:proposal2}
\end{equation}
At two loops it is equivalent to our computation, but at higher orders and at non perturbative level the equivalence does not hold anymore. This strategy is different
 from the four dimensional analogy, where $\lambda$ appeared in $a$. It would be interesting to see if even in four dimensions  there is this ambiguity.
It can just be related to a different choice of the scheme, or it may be the origin of something deeper involving the $a$-theorem. 
Indeed as stressed in \cite{Myers:2010xs,Myers:2010tj,Casini:2011kv} the quantity
respecting a c-theorem away from the  fixed point may be the entanglement entropy,  calculated for a spherical entangling surface $S_E$ from the formula 
$S_E=\log \mathcal{Z}$, where $\mathcal{Z}$ is the partition function evaluated on $S^d$.
If the entanglement entropy turns out to respect the strong version of a c-theorem, we expect that the above
ambiguity would be fixed by the requirement that $S_E=\log \mathcal{Z}$ holds at the non-perturbative level.

This analogy may lead to a  strongly coupled version of (\ref{eq:relation}), where no perturbative comparison can be made. Then, one should rely on other methods, 
as the AdS/CFT correspondence. Indeed the proposals (\ref{eq:proposal}) or (\ref{eq:proposal2}) 
can be formulated by looking at the scaling dimensions dependence on the holographic renormalization scale. On the supergravity side, the warp factor $A(r)$
plays the role of the beta function. 
At the fixed point the four dimensional central charge $a$ and the three dimensional free energy $F$ are related by the gauge gravity
correspondence to the volumes of the geometry  \cite{Martelli:2005tp,Martelli:2011qj}. 
 By using the AdS/CFT dictionary one can add the Lagrange multiplier at geometrical level
and obtain the strong coupling version of our proposal. 
Another interesting possibility is finding the four dimensional supergravity
dual of the $F$-maximization and the Lagrange multiplier extension along 
the lines of \cite{Tachikawa:2005tq}.

\section*{Acknowledgments}

We are grateful to Ken Intriligator for discussions.
A.A. is supported by UCSD grant DOE-FG03-97ER40546.  The work of M.S. is supported in part by the FWO - Vlaanderen,
Project No. G.0651.11, and in part by the Federal Office for Scientific,
Technical and Cultural Affairs through the ``Interuniversity Attraction
Poles Programme -- Belgian Science Policy'' P6/11-P.

\bibliographystyle{JHEP}
\bibliography{Lagrange}

\end{document}